\documentclass[letterpaper, 10 pt, conference]{ieeeconf}  

\IEEEoverridecommandlockouts                              
\usepackage{cite}
\usepackage{amsmath,amssymb,amsfonts}
\usepackage{algorithm,algorithmic}
\usepackage{cases}
\usepackage{graphicx}
\usepackage{textcomp}
\usepackage{dsfont}
\usepackage{enumerate}

\usepackage[caption=false,font=footnotesize]{subfig}

\usepackage{theorem}
\theorembodyfont{\rmfamily}

\newtheorem{lem}{Lemma}
\newtheorem{defin}{Definition}
\newtheorem{theorem}{Theorem}

\newtheorem{assum}{Assumption}

\renewcommand{\QED}{\QEDopen}

\newcommand{\NN}{\mathbb{N}}
\newcommand{\Zp}{{\mathbb{Z}_+}}
\newcommand{\mc}[1]{\mathcal{#1}}
\newcommand{\hs}{& \hspace{-3mm}}
\newcommand{\ssen}{s^{\rm s}}
\newcommand{\mcss}{\mathcal{S}^{\rm s}}

\newcommand{\srec}{s^{\rm r}}
\newcommand{\mcsr}{\mathcal{S}^{\rm r}}

\newcommand{\Us}{U^{\rm s}}
\newcommand{\Ur}{U^{\rm r}}
\newcommand{\iTheta}{{\it \Theta}}
\newcommand{\io}{{\rm i.o.}}
\newcommand{\tb}{\theta_{\rm b}}
\newcommand{\tm}{\theta_{\rm m}}
\newcommand{\ssk}{s^{\rm s}_k}
\newcommand{\srk}{s^{\rm r}_k}
\newcommand{\htheta}{\hat{\theta}}
\newcommand{\Exp}[1]{\mathbb{E}\left[#1\right]}

\DeclareMathOperator*{\argmax}{arg\,max}

\title{\LARGE \bf
Asymptotic Security by Model-based Incident Handlers\\ for Markov Decision Processes
}

\author{Hampei Sasahara and Henrik Sandberg
\thanks{This work was supported by Swedish Research Council.}
\thanks{H. Sasahara and H. Sandberg are with Division of Decision and Control Systems, School of Electrical Engineering and Computer Science,
        KTH Royal Institute of Technology, Stockholm SE-100 44, Sweden
        {\tt\small \{hampei,hsan\}@kth.se}}%
}

\begin{document}

\maketitle
\thispagestyle{empty}
\pagestyle{empty}

\begin{abstract}

This study investigates general model-based incident handler's asymptotic behaviors in time against cyber attacks to control systems. The attacker's and the defender's dynamic decision making is modeled as an equilibrium of a dynamic signaling game. It is shown that the defender's belief on existence of an attacker converges over time for any attacker's strategy provided that the stochastic dynamics of the control system is known to the defender. This fact implies that the rational behavior of the attacker converges to a harmless action as long as the defender possesses an effective counteraction. The obtained result supports the powerful protection capability achieved by model-based defense mechanisms.
\end{abstract}

\section{INTRODUCTION}

Secure control system design is an urgent matter as illustrated by several fatal incidents in critical infrastructures that occurred in the last decade~\cite{Nicolas2011Stuxnet,CISA2014,CISA2018,CISA2017}.
Risk assessment is necessary as one of the fundamental steps to build secure control systems.
Specifically, it is required to evaluate multiple factors, such as possibility of vulnerability, impacts of potential threats, and implementation cost for appropriate countermeasures, in a quantitative manner~\cite{NIST2012}.
In particular, counteractions carried out during adverse events are referred to as \emph{incident handling}~\cite{Paul2012Computer}.
For incident handling, which is divided into multiple steps including attack detection, influence reduction, and vulnerability elimination, we need to perform highly complicated decision making.
To handle the complexity, automatic incident handlers that utilize the dynamical model of the system to be defended have been proposed~\cite{Giraldo2018A,Xiao2018IoT}.

This study investigates behaviors of general model-based incident handlers with perfect model knowledge for risk assessment of control systems.
Mostly, a model-based incident handler passively monitors the control system's behavior, and proactively carries out a proper reaction by estimating a reasonable attack scenario if the system's behavior is inconsistent with its model.
However, even if the perfect model knowledge is available, it is impossible to choose the appropriate reaction instantaneously  owing to randomness of the environment, such as disturbance and noise.
Thus, the possibility of transient deception caused by the randomness is unavoidable to the defender.
For the sake of generality, we confine our attention to model-based incident handler's asymptotic behaviors in time.

Our main interest is to examine whether model-based incident handlers can be deceived not only transiently but also permanently.
In other words, we derive a condition under which the defender achieves appropriate reaction in a finite time step.
As a specific scenario, we suppose a powerful attacker who possesses perfect knowledge of the system model, the defender's decision making rule, and the input-output data.
For description of the system's behavior, we model the attacker's and defender's decision making as reasonable strategies of a dynamic signaling game.
Using the model, we analyze the action profiles taken with the reasonable strategies and the induced trajectories of defender's belief on the existence of the attacker.

Technically, we show that model-based incident handlers can guarantee asymptotic security as long as the defender possesses an effective counteraction.
First, it is shown that the defender's belief on the existence of the attacker converges without oscillation in a stochastic sense for any attacker's strategy.
Moreover, except for the case where the defender forms a firm belief, the control system's behavior must be consistent with the one under nominal operation for convergence of the belief.
This observation implies that attacks cannot be injected after a sufficient period of time has elapsed, without being detected.
In this sense, the control system is guaranteed to be secure in an asymptotic manner.

A number of studies have addressed security of control systems (see, e.g.,~\cite{Seyed2019Systems} and references therein).
In particular, for risk assessment, specific incident handling schemes, such as attack detection~\cite{Giraldo2018A,Gallo2020A}, resilient state estimation~\cite{Pajic2017Attack}, and attack containment~\cite{Sasahara2020DisconnectionB}, have been treated and analyzed.
Mostly, the systems have been assumed to belong to a limited class, typically linear time-invariant systems.
On the other hand, this work considers general systems and discusses a universal property of model-based incident handlers.
This generality is achieved by focusing only on asymptotic behaviors.
With respect to information system security, signaling games are often used for representing strategic and adversarial decision making~\cite{Carroll2011A,Farokhi2017Estimation,Pawlick2019Modeling,Zhu2020Secure}.
However, those works consider only one-step games, i.e., transient decision making and its influence is analyzed.
In contrast, this paper addresses asymptotic analysis of the signaling game.
Finally, this study is a generalized version of the preliminary work~\cite{Sasahara2020Asymptotic}.

This paper is organized as follows.
In Sec.~\ref{sec:model}, the system's behavior with a model-based incident handler under the supposed attack scenario is modeled as a dynamic signaling game.
In Sec.~\ref{sec:ana}, it is shown that the defender's belief of the existence of an attacker converges overt time.
This fact derives an asymptotic security of control systems when the attacker prefers to conceal her existence.
Sec.~\ref{sec:num} verifies the result through a numerical example and discusses a protection scheme based on the obtained result.
Finally, Sec.~\ref{sec:conc} draws the conclusion.


\subsection*{Notation}
Let $\NN$, $\Zp$, and $\mathbb{R}$ be the sets of natural numbers, nonnegative integers, and real numbers, respectively.
The $k$-ary Cartesian power of the set $\mc{X}$ is denoted by $\mc{X}^k.$
The filtered probability space considered in this paper is denoted by $(\Omega,\mc{F},{\rm Pr};\{\mc{F}_k\}_{k\in\NN}).$
The $\sigma$-algebra generated by a random variable $X$ is denoted by $\sigma(X)$.
The expected value of a real-valued random variable $X$ is denoted by $\mathbb{E}[X]$.
The conditional expected value of $X$ given a $\sigma$-algebra $\mc{G}$ is denoted by $\mathbb{E}[X|\mc{G}]$.
For a sequence of events $\{E_k\}_{k=1}^{\infty}\subset\mc{F},$ the supremum set $\cap_{N=1}^{\infty}\cup_{k=N}^{\infty}E_k$, namely, the event where $E_k$ occurs infinitely often, is denoted by $\{E_k\ \io\}$.
Appendix~\ref{app:proof} contains the proofs.

\section{Modeling Using Dynamic Signaling Games}
\label{sec:model}

\subsection{Motivating Example}
\label{subsec:sketch}

\begin{figure}[t]
  \centering
  \includegraphics[width=0.98\linewidth]{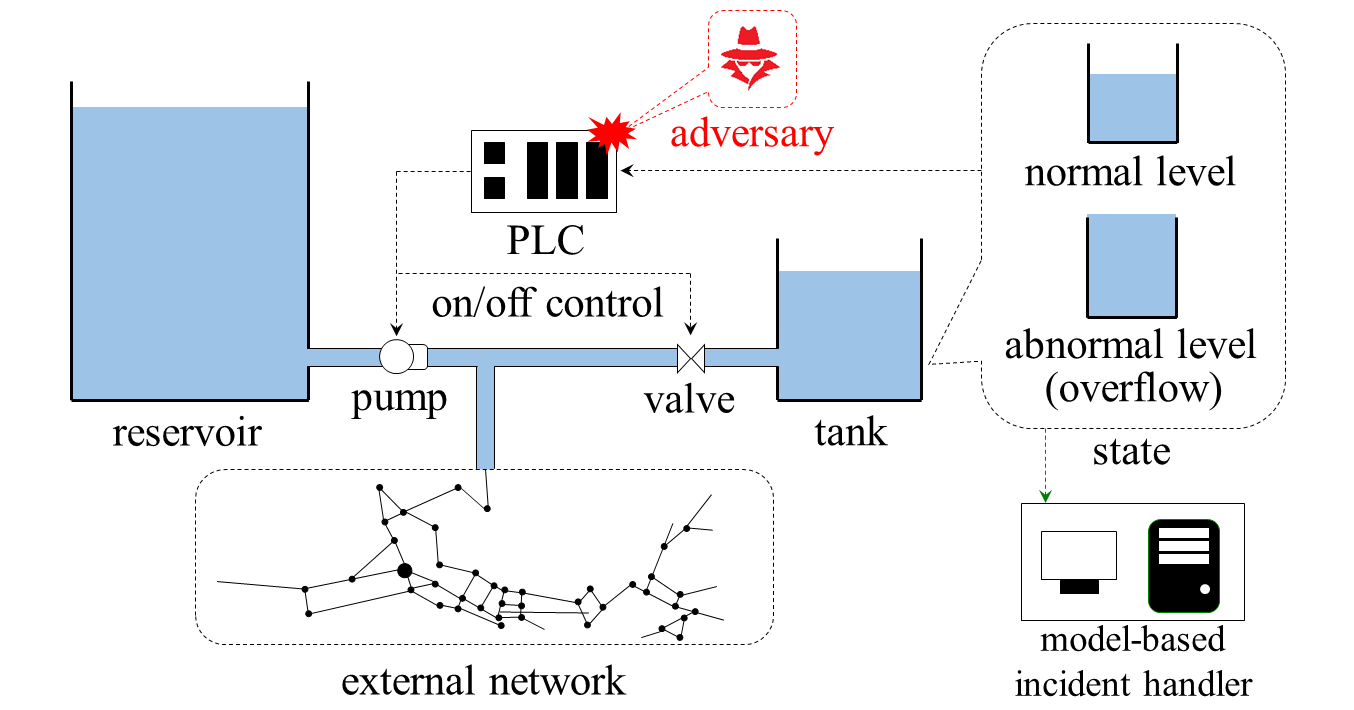}
  \caption{Motivating example: a water tank system connected to a reservoir within a water distribution network.
  The programmable logic controller (PLC) transmits on/off control signals to the pump and the valve monitoring the state, namely, the water level of the tank.
  In the supposed scenario, an adversarial software possibly intrudes into the PLC and then the infected PLC tries to cause overflow by sending inappropriate control signals without being detected.
  A model-based incident handler, which can monitor only the state, is also installed to deal with the attack.}
  \label{fig:WDN}
\end{figure}

This subsection provides a motivating example.
We here treat water distribution networks (WDNs), which supply drinking water of suitable quality to customers.
Because of their indispensability to our life, WDNs are attractive targets for adversaries~\cite{Rasekh2016Smart}.
In particular, we consider the water tank system illustrated by Fig.~\ref{fig:WDN}, where a tank is connected to a reservoir within a WDN.
The amount of the water in the tank varies due to usage for drinking and flow between the external network.
Thus the tank system is required to be properly controlled through actuation of the pump and the valve to keep the water amount within a desired range~\cite{Creaco2019Real}.
A programmable logic controller (PLC) transmits on/off control signals to the pump and the valve monitoring the state, namely, the water level of the tank.
The dynamics are modeled as a Markov decision process, where the state space and the action space are given by quantized water levels and finite control actions.
Interaction to the external network is modeled as a randomness in the process.

We here suppose an attack scenario considered in~\cite{Taormina2017Characterizing}.
The adversary succeeds to hijack the PLC and can directly manipulate its control logic.
Such an intrusion can be carried out by stealthy and evasive maneuvers in advanced persistent threats~\cite{Chen2014Study}.
The objective of the attack is to damage the system by causing water overflow through inappropriate control signals without being detected.
To deal with this attack, we suppose that a model-based incident handler, which can monitor only the state, is installed with the water tank.
The model-based incident handler chooses a proper reaction by detecting if the system is under attack through observation of the state.
If the system's behavior is highly suspicious, for example, the incident handler suggests an aggressive reaction such as log analysis or dispatch of operators.

\begin{figure}[t]
\centering
\subfloat[][Behavior of the belief on existence of an attacker.
  The belief is increased under attacks while it is expected to be decreased without attacks.]{
\includegraphics[width=.95\linewidth]{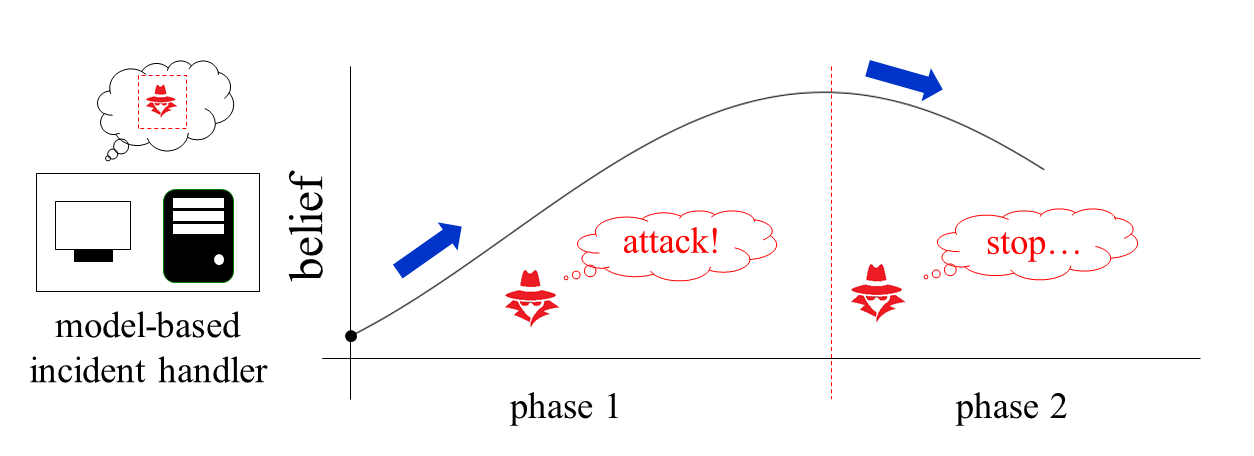}\label{subfig:confidence}
}\\
\subfloat[][Behavior of the belief when the model-based incident handler is permanently deceived.
  The belief may oscillate through intermittent attacks.]{
\includegraphics[width=.95\linewidth]{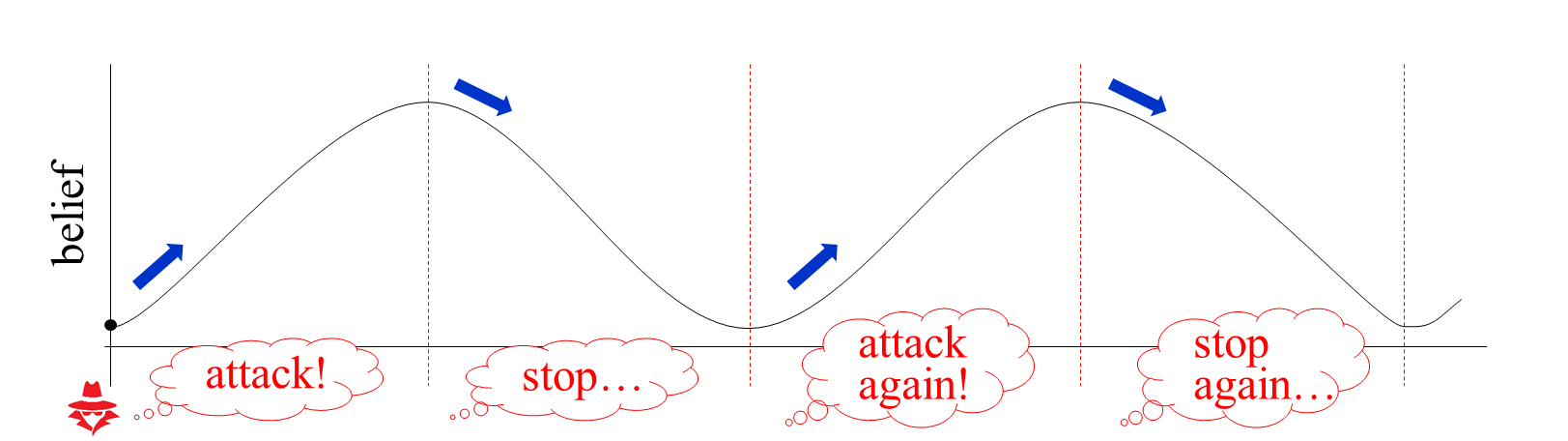}\label{subfig:confidence_perm_dec}
}
\caption{Possible behaviors of the belief on existence of an attacker.}
\end{figure}

The key notion to analyze the system's resilience is belief, namely, confidence on the existence of an attacker.
If the attacker executes an attack, then the system's behavior is different from the one under the nominal operation and accordingly the belief should be increased.
Conversely, if the attacker stays calm by choosing proper control signals, it is expected that the belief is decreased as depicted by Fig.~\ref{subfig:confidence}.
Our main interest in this study is to investigate whether the model-based incident handler is permanently deceived, i.e., the possibility of sophisticated attacks that may cause oscillation of the belief as illustrated by Fig.~\ref{subfig:confidence_perm_dec}.

\subsection{System Description}

\begin{figure}[t]
\centering
\includegraphics[width = .98\linewidth]{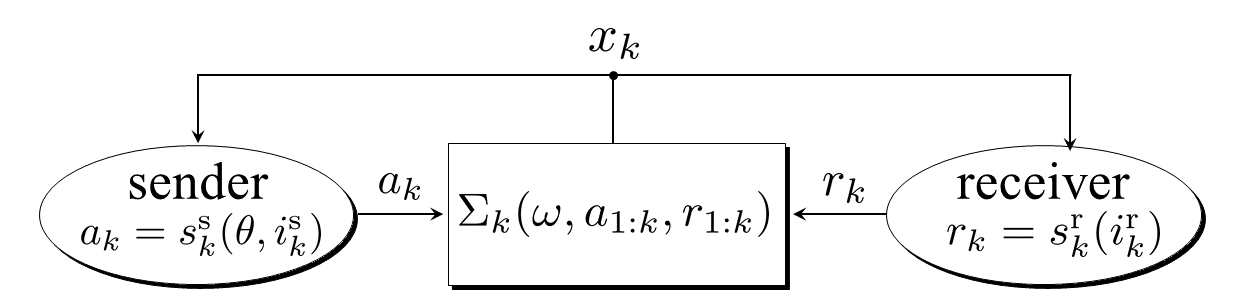}
\caption{Control system and defense architecture.
}
\label{fig:sys}
\end{figure}

Let us now introduce the general system description.
Consider a control system, possibly under attack, as depicted in Fig.~\ref{fig:sys}.
There is an agent, called a \emph{sender}, who can alter the behavior of the system through an \emph{action} $a_k\in \mc{A}$ for $k\in \NN$.
The sender can be an attacker when an adversary has intruded in the control system.
The output of the system at the $k$th step is denoted by $x_k\in \mc{X}$.
Based on the measured output, the other party, called a \emph{receiver}, chooses an action $r_k\in\mc{R}$ at each time step.
We henceforth refer to $r_k$ as a \emph{reaction} for emphasizing that $r_k$ denotes a counteraction against potentially malicious attacks.
The dynamics of the system is described by the map
\[
 \Sigma_k:\Omega\times\mc{A}^{k}\times\mc{R}^k\to\mc{X}
\] 
for $k\in \NN$ where the effect of the initial condition is disregarded for simplicity on the premise that the initial state is publicly known.
Note that, although the dynamics in the motivating example is independent of the reaction, the control system is assumed to be dependent for generality.

For simplicity, we assume that the sets of signals and actions, namely, $\mc{X},\mc{A},\mc{R}$, are finite sets.
Moreover, $\Sigma_k$ is assumed to be a time-homogeneous Markov decision process.
The transition probability from $x$ to $x'$ with $a$ and $r$ is denoted by $p(x'|x,a,r)$.
The following assumption is made to guarantee variation of the control system's behavior for different actions.
\begin{assum}\label{assum:input_obs}
For any $x\in\mc{X}$ and $r\in\mc{R}$, there exists $x'\in\mc{X}$ such that $p(x'|x,a,r)\neq p(x'|x,a',r)$
for different actions $a\neq a'$.
\end{assum}
Assumption~\ref{assum:input_obs} eliminates the possibility of stealthy attacks such as covert attack~\cite{Smith2015Covert} and zero-dynamics attack~\cite{Teixeira2012Revealing,Pasqualetti2015Control}.

Next, we formulate the decision making as a dynamic signaling game.
Let $\theta\in\iTheta$ denote the \emph{type} of the sender.
For simplicity, the type is assumed to be binary, i.e., $\iTheta=\{\tb ,\tm\},$
where $\tb$ and $\tm$ correspond to benign and malicious senders, respectively.
The types $\tb$ and $\tm$ describe the situations where there does not and do exist an adversary in the control system, respectively.
This binary assumption implies that we focus on a single threat scenario.
For handling multiple scenarios, it suffices to consider multiple-valued types.
Let $\{\ssk\}_{k\in\NN}$ and $\{\srk\}_{k\in\NN}$ denote the sender's and receiver's \emph{strategy} profiles, respectively.
The strategies at the $k$th step are given by
\[
 \ssk:\iTheta\times\mc{I}^{\rm s}_k\to\mc{A},\quad \srk:\mc{I}^{\rm r}_k\to\mc{R}
\]
where $i^{\rm s}_k\in\mc{I}^{\rm s}_k$ and $i^{\rm r}_k\in\mc{I}^{\rm r}_k$ are information sets at the $k$th step given by
\[
 i^{\rm s}_k=(x_{1:k-1},a_{1:k-1}),\quad i^{\rm r}_k=(x_{1:k-1},r_{1:k-1}).
\]
These information sets imply measurability of the state and perfect recall of the agents' decisions.
We hereinafter denote the strategy profile of each player by $s^{\rm s}:=\{\ssk\}_{k\in\NN}$, $s^{\rm r}:=\{\srk\}_{k\in\NN}$
and the pair of them by $s:=(s^{\rm s},s^{\rm r}).$
The sender's and receiver's admissible strategy sets are denoted by $\mcss$ and $\mcsr$, respectively.

In preparation for the game-theoretic formulation in the sequel,
\begin{equation}\label{eq:signals}
\begin{array}{ll}
 X^{\theta,s}_k(\omega) \hs := \Sigma_k(\omega,A^{\theta,s}_{1:k}(\omega),R^{\theta,s}_{1:k}(\omega)),\\
 A^{\theta,s}_k(\omega) \hs := s^{\rm s}_k(\theta,I^{{\rm s},\theta,s}_k(\omega)),\quad
 R^{\theta,s}_k(\omega) := s^{\rm r}_k(I^{{\rm r},\theta, s}_k(\omega))
\end{array}
\end{equation}
where
\[
 \begin{array}{ll}
  I^{{\rm s},\theta,s}_k(\omega) \hs := (X^{\theta,s}_{1:k-1}(\omega), s^{\rm s}_{1:k-1} (\theta,I^{{\rm s},\theta,s}_{1:k-1}(\omega))),\\
  I^{{\rm r},\theta,s}_k(\omega) \hs := (X^{\theta,s}_{1:k-1}(\omega), s^{\rm r}_{1:k-1} (I^{{\rm r},\theta,s}_{1:k-1}(\omega))).
 \end{array}
\]
We denote the conditional probability mass function of $X^{\theta,{\rm s}}_{k+1}$ given $x_{1:k}$ by $p^{\theta,s}_{k+1}(x_{k+1}|x_{1:k})$.

This setup describes the situation where the incident handler does not know whether the control system is attacked, or not.
This game is thus categorized into the class of incomplete information games, and in particular, signaling games because the type of a player is unknown to the opponent.

\subsection{Signaling Game Setup}

To define reasonable strategies, we introduce \emph{belief systems}, for the receiver, on the sender's type.
A belief system is a tuple of the functions $\pi_k:\iTheta\times\mc{X}^k\to[0,1].$
As the belief is close to one, the receiver believes that the sender is malicious with high confidence.
When the following conditions are satisfied, the belief system $\pi:=\{\pi_k\}_{k\in\Zp}$ is said to be consistent with the strategy profile $s$:
\begin{itemize}
\item The initial belief satisfies $\pi_0(\htheta)\geq 0$ for any $\htheta\in\iTheta$ and $\sum_{\htheta\in\iTheta}\pi_0(\htheta)=1$.
\item For any $k\in\Zp$ and $x_{1:k+1}\in\mc{X}^{k+1}$ that satisfy the condition
$\sum_{\phi\in\iTheta}p^{\phi,s}_{k+1}(x_{k+1}|x_{1:k})\pi_k(\phi;x_{1:k})\neq0$,
the transition follows Bayes' rule determined by $s$:
\[
\begin{array}{l}
 \pi_{k+1}(\htheta;x_{1:k+1})=f^s_{k+1}(\htheta,x_{1:k+1})\pi_k(\htheta;x_{1:k})
\end{array}
\]
where
\[
 f^s_{k+1}(\htheta,x_{1:k+1}) := \dfrac{p^{\htheta,s}_{k+1}(x_{k+1}|x_{1:k})}{\sum_{\phi\in\iTheta}p^{\phi,s}_{k+1}(x_{k+1}|x_{1:k})\pi_k(\phi;x_{1:k})}.
\]
\end{itemize}
The initial belief $\pi_0$ is assumed to be known to both players.
Note that $\htheta$ represents not the true type but an estimated type by the receiver in the notation.

As a decision rule for rational players' strategies, we consider uniform equilibria.
Let the sender's instantaneous utility be given by $U^{\rm s}:\iTheta\times\mc{X}\times\mc{A}\times\mc{R}\to\mathbb{R}$.
The sender's expected average utility up to the $T$th step is given by
\[
 \bar{U}^{\rm s}_T(\theta,s):= \Exp{\dfrac{1}{T} \sum_{k=1}^{T} U^{\rm s}(\theta,X^{\theta,s}_k,A^{\theta,s}_k,R^{\theta,s}_k)}.
\]
Similarly, with the receiver's instantaneous utility given by $U^{\rm r}:\mc{X}\times\mc{A}\times\mc{R}\to\mathbb{R}$,
the receiver's expected average utility up to the $T$th step is given by
\[
 \begin{array}{l}
 \bar{U}^{{\rm r}}_T(s,\pi):= \\
 \displaystyle{ \Exp{\dfrac{1}{T} \sum_{k=1}^{T} \sum_{\htheta\in\iTheta}U^{\rm r}(\htheta,X^{\htheta,s}_k,A^{\htheta,s}_k,R^{\htheta,s}_k) \pi_k(\htheta;X^{\htheta,s}_{1:k})}}.
\end{array}
\]
Under this notation, the strategy profile $s=(\ssen,\srec)$ is said to be a \emph{Bayesian-Nash equilibrium} if $(\bar{U}^{\rm s}_T(\theta,s),\bar{U}^{{\rm r}}_T(s,\pi))$ converges to $(\bar{U}^{{\rm s}}(\theta,s),\bar{U}^{{\rm r}}(s,\pi))$
as $T\to \infty$
with a consistent belief system $\pi$ and
\[
\left\{
\begin{array}{l}
 \displaystyle{\ssen \in \argmax_{\hat{s}^{\rm s} \in \mcss}\ \bar{U}^{{\rm s}}(\theta,(\hat{s}^{\rm s},\srec))}\quad \forall \theta\in\iTheta,\\
 \displaystyle{\srec \in \argmax_{\hat{s}^{\rm r} \in \mcsr}\ \bar{U}^{{\rm r}}((\ssen,\hat{s}^{\rm r}),\pi)}
\end{array}
\right.
\]
is satisfied.
Note that the results in this paper can be extended to the case when the utility is not taken to be uniform, such as discounted utilities.

The subsequent section analyzes properties of reasonable strategies on the premise that an equilibrium exists although its existence is a fundamental issue to be investigated.
It is known that mixed strategies admit existence of equilibria in most cases.
For example, repeated games with complete information always have a Nash equilibrium in mixed strategies~\cite[Chap.~8]{Laraki2019Mathematical}.
Although we consider only pure strategies in this paper, an extension to mixed strategies is straightforward.

\section{Analysis}
\label{sec:ana}

In this section, we analyze asymptotic behaviors of beliefs and actions for detection-averse strategies.
It is shown that the control system is guaranteed to be secure in an asymptotic manner as long as the defender possesses an effective counteraction.
Throughout this section, we fix a strategy profile given as an equilibrium and omit $s$ in the notation for simplicity.

\subsection{Belief's Asymptotic Behavior}
\label{subsec:conv}

First, we investigate asymptotic behaviors of beliefs.
Suppose that a true type $\theta$, a strategy profile $s$ and a belief system $\pi$ are given as an equilibrium.
Then the value of the belief about $\htheta$ at the $k$th step for each outcome $\omega\in\Omega$ is given by
\[
 \pi^{\htheta,\theta}_k(\omega):=\pi_k(\htheta;X^{\theta}_{1:k}(\omega)).
\]
We denote the belief on the true type by
$\pi^{\htheta=\theta}_k(\omega) := \pi^{\theta,\theta}_k(\omega)$
in the following discussion.

First of all, the following key lemma holds.
\begin{lem}\label{lem:submar}
For any type $\theta$, strategy profile $s$, and consistent belief system $\pi$,
the belief on the true type
 $\pi^{\htheta=\theta}_k$
is a submartingale with respect to the filteration
$\sigma(X^{\theta}_{1:k})$.
\end{lem}
Lemma~\ref{lem:submar} implies that the belief on the true type is non-decreasing in a stochastic sense.
As a direct conclusion of this lemma, obtained by the Doob's convergence theorem, the following theorem holds.
\begin{theorem}\label{thm:conv}
For any type $\theta$, strategy profile $s$, and consistent belief system $\pi$,
the belief on the true type
$\pi^{\htheta=\theta}_k$
converges almost surely as $k\to \infty$.
\end{theorem}
Theorem~\ref{thm:conv} implies that the belief does not oscillate even under an intermittent attack as in Fig.~\ref{subfig:confidence_perm_dec}.
We denote the limit by
$\pi^{\htheta=\theta}_{\infty}:\Omega\to[0,1]$
where
\[
  \pi^{\htheta=\theta}_k \to \pi^{\htheta=\theta}_{\infty}\quad {\rm a.s.}
\]
as $k\to \infty$.


{\it Remark:}
A heuristic justification of Theorem~\ref{thm:conv} from an information-theoretic perspective can be given as follows.
Suppose that the true type is $\tm$ and the state sequence $x_{1:k}$ is observed.
Then the belief is given by
\begin{equation}\label{eq:pi_KL}
 \begin{array}{cl}
 \pi^{\htheta=\tm}_k
 \hs = \dfrac{\pi_0(\tm)}
 {(p^{\tb}(x_{1:k})/p^{\tm}(x_{1:k}))\pi_0(\tb) + \pi_0(\tm)}\\
 \hs = \dfrac{\pi_0(\tm)}
 {{\rm exp}(kS_k) \pi_0(\tb) + \pi_0(\tm)}
 \end{array}
\end{equation}
where
$p^{\theta}(x_{1:k})$
is the joint probability mass function of $x_{1:k}$ and
\[
 S_k:=\dfrac{1}{k} \sum_{i=1}^k \log \dfrac{p^{\tb}(x_i|x_{1:i-1}) }
 {p^{\tm}(x_i|x_{1:i-1})}.
\]
Assuming that
$p^{\theta}_k(x_k|x_{1:k-1})$
approaches a stationary distribution
$p^{\theta}$
and the strong law of large numbers (SLLN) can be applied, for sufficiently large $k$ we have
\[
 S_k\simeq\mathbb{E}_{x\sim p^{\tm}}
 \left[
 \log p^{\tb}(x)/p^{\tm}(x)
 \right]
 = -D_{\rm KL}(p^{\tm}||p^{\tb})
\]
where $D_{\rm KL}$ denotes the Kullback-Leibler divergence.
Since $D_{\rm KL}$ is nonnegative for any pair of distributions, $S_k$ converges to a nonnegative number, which results in convergence of
$\pi^{\htheta=\tm}_k$.
Bayesian estimator's convergence to the true parameter, referred to as Bayesian consistency, has been investigated mainly in the context of statistics~\cite{Diaconis1986On}.
In this sense, Theorem~\ref{thm:conv} can be regarded as another representation of Bayesian consistency in the context of security.
However, note again that this discussion is not a rigorous proof but a heuristic explanation since the state is essentially non-i.i.d. (independent and identically distributed) and applicability of SLLN cannot be ensured.

\subsection{Definition of Detection-averse Utilities}
\label{subsec:det_sens_util}

To clarify our interest, we define the notion of detection-averse utilities.
\begin{defin}\label{defin:det_sens_util}
{\bf (Detection-averse Utilities)}
A pair $(\Us,\Ur)$ is said to be detection-averse utilities when 
\begin{equation}\label{eq:det_sens_str}
 \pi^{\htheta=\tm}_\infty < 1\quad {\rm a.s.}
\end{equation}
for any Bayesian-Nash equilibrium $s$ and consistent belief system $\pi$.
\end{defin}
Definition~\ref{defin:det_sens_util} characterizes utilities with which the malicious sender avoids having the defender form a firm belief on the existence of an attacker.
In other words, the reasonable strategy becomes detection-averse when the defender possesses an effective counteraction.
If the utilities of interest are not detection-averse, there is no trade-off from the attacker's perspective.
For protecting such systems, design of appropriate counteractions should be performed as a premise of the presented framework.

Examples of effective counteractions include \emph{fallback control}~\cite{Sasaki2015Model} and \emph{separation-based reconfiguration}~\cite{Sasahara2020DisconnectionB}.
Suppose that the control system to be protected is networked and connected to the Internet through, for example, human machine interface.
Those proposed methods detect an unauthorized access and exclude the attacker by disconnecting the attacked components.
When the attacker prefers to lurk without being detected and keep the unauthorized access, this situation can be modeled with detection-averse utilities.

\subsection{Asymptotic Security}
\label{subsec:asym_sec}

As a preparation of our main claim, we investigate the asymptotic behavior of state transition.
By dividing the cases with respect to the limit of the belief, we obtain the following lemma.
\begin{lem}\label{lem:case}
For any type $\theta$, strategy profile $s$, and consistent belief system $\pi$,
we have
\[
 {\rm Pr}( E^{\theta}_{f\to 1} \cup E^{\theta}_{\pi\to 0})=1
\]
where
$E^{\theta}_{f\to 1}$
is the event where the coefficient of Bayes' rule converges to one and
$E^{\theta}_{\pi \to 0}$
is the event where the belief converges to zero, i.e.,
\[
\begin{array}{l}
 E^{\theta}_{f\to 1}:=\{\omega\in\Omega: f^{\htheta=\theta}_k(\omega)\to 1\},\\
 E^{\theta}_{\pi \to 0}:=\{\omega\in\Omega: \pi^{\htheta=\theta}_{\infty}(\omega)=0\}
\end{array}
\]
with
\[
 f^{\htheta=\theta}_k(\omega):= f_k(\theta,X^{\theta}_{1:k}(\omega)),
\]
which is the coefficient in Bayes' rule for the true type.
\end{lem}
Lemma~\ref{lem:case} implies that there are only two cases: one is that the belief update gradually stops and the other is that the belief on the true type converges to zero.

We here need a technical assumption to eliminate the latter case.
\begin{assum}\label{assum:zero_zero}
For any type $\theta$, strategy profile $s$, and consistent belief system $\pi$,
${\rm Pr}(E^{\theta}_{\pi \to 0})=0$
holds.
\end{assum}
Assumption~\ref{assum:zero_zero} guarantees that the belief on the true type does not converge to zero.
A control system that satisfies Assumption~\ref{assum:zero_zero} is provided in Appendix~\ref{app:ex}.

Under Assumption~\ref{assum:zero_zero}, Lemma~\ref{lem:case} implies that the coefficient of Bayes' rule converges to one almost surely.
This claim is equivalent to that the state eventually loses information on the type.
\begin{lem}\label{lem:pp}
Let Assumption~\ref{assum:zero_zero} hold.
Every Bayesian-Nash equilibrium with detection-averse utilities satisfies
\[
 |p^{\htheta_{\rm b},\tm}_k-p^{\htheta_{\rm m},\tm}_k| \to 0\quad {\rm a.s.}
\]
where
\[
\begin{array}{l}
 p^{\htheta_{\rm b},\tm}_k(\omega):=p^{\tb}_k(X_k^{\tm}(\omega)|X_{1:k-1}^{\tm}(\omega)),\\
 p^{\htheta_{\rm m},\tm}_k(\omega):=p^{\tm}_k(X_k^{\tm}(\omega)|X_{1:k-1}^{\tm}(\omega)).
\end{array}
\]
\end{lem}
For interpretation of Lemma~\ref{lem:pp}, consider the ideal case where
$p^{\htheta_{\rm b},\tm}_k=p^{\htheta_{\rm m},\tm}_k$
holds at some time step $k$.
This condition means that the transition of the state's probability mass function is identical regardless of the estimated type.
In other words, the state does not possess information about the attacker's type.
Therefore, we can interpret Lemma~\ref{lem:pp} as the fact that the state has to lose information on the type asymptotically.

From Lemma~\ref{lem:pp} and Assumption~\ref{assum:input_obs}, the actions themselves must be identical.
This fact yields the main result of this study: asymptotic security is achieved by model-based incident handlers.
\begin{theorem}\label{thm:asymptotic_security}
Let Assumptions~\ref{assum:input_obs} and~\ref{assum:zero_zero} hold.
Every Bayesian-Nash equilibrium with detection-averse utilities satisfies
\[
 d(A^{\htheta_{\rm b},\tm}_k,A^{\htheta_{\rm m},\tm}_k)\to 0\quad {\rm a.s.}
\]
with $ A^{\htheta_{\rm b},\tm}_k(\omega):= s^{\rm s}_k(\tb,I^{{\rm s},\tm}(\omega))$ and $A^{\htheta_{\rm m},\tm}_k(\omega):= s^{\rm s}_k(\tm,I^{{\rm s},\tm}(\omega))$
where $d:\mc{A}\times\mc{A}\to[0,\infty)$ is a distance given by
\[
 d(a,a')=\left\{
 \begin{array}{ll}
 0 & {\rm if}\ a=a',\\
 1 & {\rm otherwise},
 \end{array}
 \right.
\]
which induces the discrete topology.
\end{theorem}
Theorem~\ref{thm:asymptotic_security} implies that the malicious sender's action converges to the benign one.
Equivalently, an attacker necessarily behaves as a benign sender after a sufficiently large step.
Therefore, the control system is guaranteed to be secure in an asymptotic manner, i.e., model-based incident handlers are {\it never} deceived permanently.
This result indicates the powerful defense capability achieved by model knowledge.

\section{Numerical Example and Discussion}
\label{sec:num}

\subsection{Numerical Example}
\begin{figure}[t]
  \centering
  \includegraphics[width=0.98\linewidth]{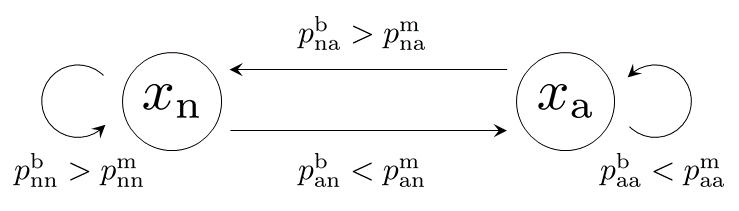}
  \caption{Example of state transition diagram with binary state and action spaces.}
  \label{fig:mot_MDP}
\end{figure}

We confirm the theoretical results through numerical simulation.
We assume the state space and the action space to be binary, i.e.,
$\mc{X} = \{x_{\rm n}, x_{\rm a}\}$ and $\mc{A} = \{a_{\rm b},a_{\rm m}\}.$
The states $x_{\rm n}$ and $x_{\rm a}$ represent the normal and abnormal states, respectively,
and $a_{\rm b}$ and $a_{\rm m}$ represent benign and malicious actions, respectively.
The benign and malicious actions correspond to proper and improper control signals, respectively.
The reaction set is given by $\mc{R}=\{r_{\rm b},r_{\rm m}\}$.
As in the motivating example, we assume that the transition probability is independent of the reaction.
The state transition diagram is depicted by Fig.~\ref{fig:mot_MDP}, where the transition probability from $x_{\rm n}$ to $x_{\rm a}$ with $a_{\rm b}$ is denoted by $p^{\rm b}_{\rm an}$, and the other transition probabilities are denoted in a similar manner.
The inequalities in Fig.~\ref{fig:mot_MDP} mean that the malicious action leads to a higher probability of the abnormal state than the benign action.
The specific values of the transition probabilities are given in Table~\ref{table:prob}, where each value corresponds to the probability from the state in the row to the state in the column.
The utilities are given in Table~\ref{table:util}, which implies that the benign sender always prefers the normal state,
the receiver always prefers the reaction corresponding to the true type,
the malicious sender prefers non-aggressive reaction, and also the abnormal state for non-aggressive reaction.
The initial state is $x_{\rm n}$.
The initial belief is given by $\pi_0(\tm)=0.1$.

\begin{table}[t]
\centering
\caption{Transition Probabilities. left: probabilities with the benign action. right: probabilities with the malicious action.}
\begin{tabular}{c|cc}
 $a_{\rm b}$ & $x_{\rm n}$ & $x_{\rm a}$\\ \hline
 $x_{\rm n}$ & 0.9 & 0.1 \\
 $x_{\rm a}$ & 0.8 & 0.2
\end{tabular}
\hspace{3mm}
\begin{tabular}{c|cc}
 $a_{\rm m}$ & $x_{\rm n}$ & $x_{\rm a}$\\ \hline
 $x_{\rm n}$ & 0.8 & 0.2 \\
 $x_{\rm a}$ & 0.7 & 0.3
\end{tabular}
\label{table:prob}
\end{table}

Since it is difficult to compute an exact equilibrium for the infinite time horizon problem,
we consider a sequence of equilibria for a finite time horizon problem.
Define the finite time horizon average utilities by
\[
 \bar{U}^{\rm s}_{k,T}:= \Exp{\dfrac{1}{T} \sum_{i=k}^{k+T-1} U^{\rm s}(\theta,X^{\theta,s}_i,A^{\theta,s}_i,R^{\theta,s}_i)}.
\]
and
\[
 \begin{array}{l}
 \bar{U}^{{\rm r}}_{k,T}(s,\pi):= \\
 \displaystyle{ \Exp{\dfrac{1}{T} \sum_{i=k}^{k+T-1} \sum_{\htheta\in\iTheta}U^{\rm r}(\htheta,X^{\htheta,s}_i,A^{\htheta,s}_i,R^{\htheta,s}_i) \pi_i(\htheta;X^{\htheta,s}_{1:i})}}.
\end{array}
\]
With those utilities, the obtained $\ssen_k$ and $\srec_k$ are used for the $k$th strategy, in a manner similar to receding horizon control.
The horizon length is given by $T=2$.

\begin{table}[t]
\centering
\caption{Utilities}
\begin{tabular}{c|cc}
 $U^{\rm s}_0(\tb)$ & $r_{\rm b}$ & $r_{\rm m}$\\ \hline
 $x_{\rm n}$ & 1 & 1 \\
 $x_{\rm a}$ & 0 & 0
\end{tabular}
\hspace{1mm}
\begin{tabular}{c|cc}
 $U^{\rm s}_0(\tm)$ &  $r_{\rm b}$ & $r_{\rm m}$\\ \hline
 $x_{\rm n}$ & 1 & 0 \\
 $x_{\rm a}$ & 2 & 0
\end{tabular}
\hspace{1mm}
\begin{tabular}{c|cc}
 $U^{\rm r}_0$ &  $r_{\rm b}$ & $r_{\rm m}$\\ \hline
 $\tb$ & 1 & 0 \\
 $\tm$ & 0 & 1
\end{tabular}
\label{table:util}
\end{table}

\begin{figure}[t]
  \centering
  \includegraphics[width=0.98\linewidth]{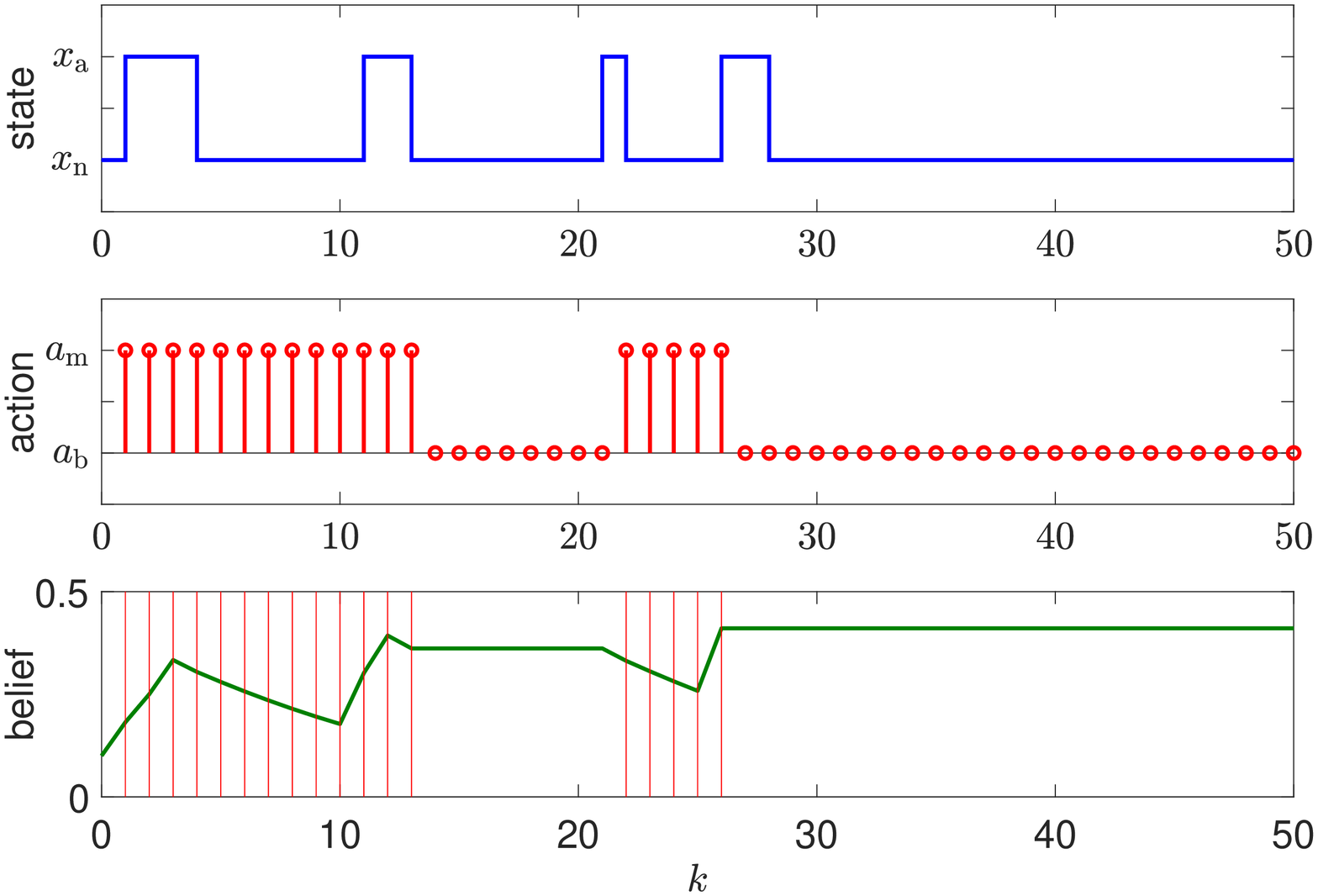}
  \caption{Sample paths of the state, the action, and the belief when $\theta=\tm$.}
  \label{graph:short}
\end{figure}

Under this setting, a sample path of the state, the action, and the belief on $\tm$ for $\theta=\tm$ is depicted in Fig.~\ref{graph:short}.
The vertical lines in the graph of belief means the action at the time instant is $a_{\rm m}$.
For $0\leq k \leq 12$, the belief is sufficiently small, and thus $a_{\rm m}$ is the rational action.
For $13\leq k <21$, the belief is large, and hence $a_{\rm b}$ is taken when the state is $x_{\rm n}$.
At $k=21$, the state is $x_{\rm a}$, and the belief begins to decrease.
At $k=26$, the state is $x_{\rm n}$.
Then the belief exceeds the threshold, and $a_{\rm b}$ regardless of the state is the rational action.
This result coincides with Theorem~\ref{thm:asymptotic_security}.
Note that, the reaction is always $r_{\rm b}$ when the belief is less than $0.5$, and hence the sender's instantaneous utility depends only on the state and the receiver's utility is one in this example.

\begin{figure}[t]
  \centering
  \includegraphics[width=0.98\linewidth]{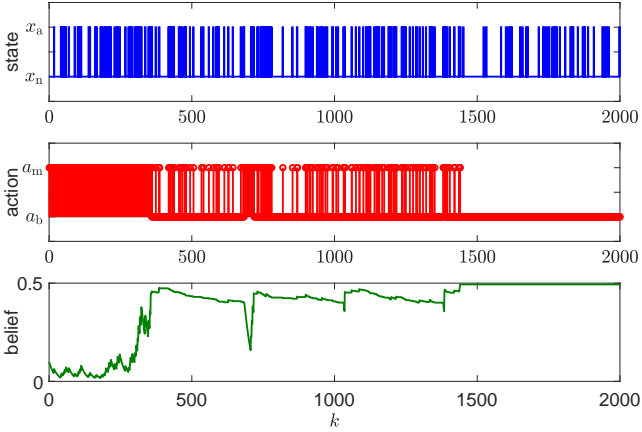}
  \caption{Sample paths of the state, the action, and the belief when $\theta=\tm$ and the transition probability is given by Table~\ref{table:prob2}.}
  \label{graph:long}
\end{figure}

To investigate a long-term behavior, consider a situation where detection is more difficult.
Specifically, the transition probability is given in Table~\ref{table:prob2}, which means the deviation of the transition probability by $a_{\rm m}$ is small.
A sample path of the state, the action, and the belief on $\tm$ for $\theta=\tm$ is depicted in Fig.~\ref{graph:long}.
Although the convergence speed is later than Fig.~\ref{graph:short}, the asymptotic security claim in Theorem~\ref{thm:asymptotic_security} can be confirmed.

\begin{table}[t]
\centering
\caption{Transition Probabilities. left: probabilities with the benign action. right: probabilities with the malicious action.}
\begin{tabular}{c|cc}
 $p^{\rm b}$ & $x_{\rm n}$ & $x_{\rm a}$\\ \hline
 $x_{\rm n}$ & 0.9 & 0.1 \\
 $x_{\rm a}$ & 0.8 & 0.2
\end{tabular}
\hspace{3mm}
\begin{tabular}{c|cc}
 $p^{\rm m}$ & $x_{\rm n}$ & $x_{\rm a}$\\ \hline
 $x_{\rm n}$ & 0.85 & 0.15 \\
 $x_{\rm a}$ & 0.79 & 0.21
\end{tabular}
\label{table:prob2}
\end{table}

\subsection{Discussion: Protection by Passive Bluffing}
\label{subsec:dis}

Roughly speaking, the result in Section~\ref{sec:ana} claims that the defender always wins in an asymptotic manner when the stochastic model of the control system is completely known and \emph{the vulnerability is known and modeled.}
The latter requirement is quantitatively described by the condition $\pi_0(\tm)>0$.
Although the derived result claims a quite powerful defense capability, it is also true that it is almost impossible to be aware of all possible vulnerabilities in advance and to prepare appropriate counteraction for all scenarios.

As a practically interesting defense scheme, it may be possible to use the obtained property for \emph{passive bluffing}.
Suppose that the attacker does not know whether her attack scenario is supposed $(\pi_0(\tm)>0)$ or not $(\pi_0(\tm)=0)$.
Also, imposing a certain property into the control system, we assume that state observation does not provide information about the reaction.
For instance, the control system in the numerical example, where the behavior is independent of the reaction, satisfies this property.
Under those assumptions, if the defender can conceal the actually conducted reactions, the true belief is completely unknown to the attacker.
In this case, even if the attack is actually a zero-day attack through an unknown vulnerability $(\pi_0(\tm)=0)$, there is a possibility to be able to protect the control system.
Specifically, if the attacker is risk-averse, i.e., she cares about the case $\pi_0(\tm)>0$, then she would possibly stop the attack after a while in a rational manner although the attack is unnoticed.
Analysis of such passive bluffing utilizing the powerful detection capability achieved by model-based incident handling is a possible future direction.

\section{Conclusion}
\label{sec:conc}

This study has investigated behaviors of model-based incident handlers using the framework of dynamic signaling games.
It has been shown that the control system can be guaranteed to be secure in an asymptotic manner when the defender possesses an effective counteraction.
Future work includes generalization of the results and a formal analysis of passive bluffing discussed in Sec.~\ref{subsec:dis}.


\appendices

\section{Example Ensuring Assumption~\ref{assum:zero_zero}}
\label{app:ex}
This appendix provides a simple example of a system that ensures the condition of Assumption~\ref{assum:zero_zero}.
Consider a binary state space and assume all transition probabilities are uniformly set to $1/2$ at the equilibrium when $\theta=\tm$. 
Assume also that the transition probabilities from one state to the other are $p\neq 1/2$ when $\theta=\tb$.
Define $E_{2k}$ as the event that the number of reaching one state is equal to the number of reaching the other state at the time step $2k$.
From the random walk theory, $E_{2k}$ occurs infinitely often almost surely.
If $\omega\in E_{2k}$, then the belief at the $2k$th step is given by $\pi^{\htheta,\theta}_{2k}(\omega) = \pi_0(\tm)/(\alpha(1-\pi_0(\tm))+\pi_0(\tm))$
with $\alpha=p^k(1-p)^k4^k$.
Because $0<\alpha<1$, we have $\pi^{\htheta=\theta}_{2k}(\omega)>\pi_0(\tm)$.
Since $E_{2k}$ occurs infinitely often almost surely, the condition of Assumption~\ref{assum:zero_zero} holds.
It is expected that a similar justification can be applied to a broader class of systems.

\section{Proofs}
\label{app:proof}
\begin{proof}
\emph{Proof of Lemma~\ref{lem:submar}:}
Since it is clear that the belief is adapted to the filteration and integrable, it suffices to show
\[
 \Exp{\pi^{\htheta=\theta,s}_{k+1}|\sigma(X^{\theta,s}_{1:k})}\geq \pi^{\htheta=\theta,s}_k\quad {\rm a.s.}
\]
for the claim.
Fix $\omega\in\Omega$ and denote $X^{\theta,s}_{1:k}(\omega)$ by $x_{1:k}$.
Then the inequality is equivalent to
\begin{equation}\label{eq:submar}
 \sum_{x_{k+1}\in\mc{X}} p^{\theta,s}_{k+1}(x_{k+1}|x_{1:k}) \pi_{k+1}(\htheta;x_{1:k+1})\geq \pi_k(\htheta;x_{1:k})
\end{equation}
for $\htheta=\theta$.
Thus it suffices to show~\eqref{eq:submar} for any $k\in\NN$ and $x_{1:k}\in\mc{X}^k$.

First, we reduce the index of the summation in~\eqref{eq:submar}.
When $\pi_k(\htheta;x_{1:k})=0$, the inequality~\eqref{eq:submar} always holds.
Thus it is assumed that $\pi_k(\htheta;x_{1:k})>0$ in the following.
Define
\[
 \mc{X}^0_k:=\left\{x_{k+1}\in\mc{X}: \sum_{\htheta\in\iTheta}p^{\htheta,s}_{k+1}(x_{k+1}|x_{1:k})\pi_k(\htheta;x_{1:k})=0\right\}.
\]
Because $\pi_k(\htheta;x_{1:k})$ is positive, if $x_{k+1}$ belongs to $\mc{X}^0_k$ then $p^{\htheta,s}_{k+1}(x_{k+1}|x_{1:k})=0$ holds.
Hence~\eqref{eq:submar} is equivalent to
\begin{equation}\label{eq:submar2}
 \sum_{x_{k+1}\in\mc{X}^+_k} p^{\theta,s}_{k+1}(x_{k+1}|x_{1:k}) \pi_{k+1}(\htheta;x_{1:k+1}) \geq \pi_k(\htheta;x_{1:k})
\end{equation}
where $\mc{X}^+_k:=\mc{X}\setminus\mc{X}^0_k$.

For notational simplicity, we define
\[
 \pi(\htheta):=\pi_k(\htheta;x_{1:k}),\ p^{\theta}(x):=p^{\theta,s}(x|x_{1:k}),\ \mc{X}^+:=\mc{X}^+_k
\]
for fixed $k$ and $x_{1:k}$.
Under this notation, since $\pi$ is consistent with $s$, the inequality~\eqref{eq:submar2} is equivalent to
\[
 \sum_{x\in\mc{X}^+} p^{\theta}(x)
 \dfrac{p^{\htheta}(x) \pi(\htheta) }
 {\sum_{\phi\in\iTheta} p^\phi(x) \pi(\phi) }
 \geq \pi(\htheta).
\]
Because $\htheta=\theta$ and $\pi(\theta)>0$, this inequality is equivalent to
\begin{equation}\label{eq:submar3}
 \underbrace{
 \sum_{x\in\mc{X}^+}
 p^{\theta}(x)
 \dfrac{ p^{\theta}(x)}
 {\sum_{\phi\in\iTheta} p^\phi(x) \pi(\phi) }
 }_{=:G(\theta)}
 \geq 1.
\end{equation}
By rewriting the left-hand side and applying Jensen's inequality, we have
\[
\begin{array}{cl}
 G(\theta) \hs
 = 
 \displaystyle{\sum_{x\in\mc{X}^+}}
 \dfrac{p^{\theta}(x)}
 {\pi(\theta)+p^{\theta'}(x)/p^{\theta}(x)\pi(\theta')}\\
 \hs  \geq
 \dfrac{1}
 {\pi(\theta)+
  \sum_{x\in\mc{X}^+}
 p^{\theta}(x)
 p^{\theta'}(x)/p^{\theta}(x)\pi(\theta')}\\
 \hs \geq
  \dfrac{1}{\pi(\theta)+\pi(\theta')}\\
 \hs = 1,
\end{array}
\]
which leads to the claim.
\end{proof}

\vspace{3mm}

\begin{proof}
\emph{Proof of Theorem~\ref{thm:conv}:}
Because the belief is uniformly bounded, we have $\sup_{k\in\NN} \Exp{\pi^{\htheta=\theta,s}_k}<\infty.$
From Lemma~\ref{lem:submar} and Doob's convergence theorem~\cite[Theorem~4.1]{Cinlar2011Probability}, the claim holds.
\end{proof}

\vspace{3mm}

\begin{proof}
\emph{Proof of Lemma~\ref{lem:case}:}
Assume $\pi^{\htheta=\theta,s}_{\infty}(\omega)=\alpha\in(0,1]$.
Then we have
\[
 \begin{array}{cl}
  \displaystyle{ \lim_{k\to\infty} f^{\theta,s}_k(\omega)} \hs \displaystyle{ =\lim_{k\to\infty} \pi^{\htheta=\theta,s}_{k+1}(\omega)/\pi^{\htheta=\theta,s}_k(\omega)}  = \alpha/\alpha = 1,
 \end{array}
\]
which leads to the claim.
\end{proof}

\vspace{3mm}

\begin{proof}
\emph{Proof of Lemma~\ref{lem:pp}:}
From Lemma~\ref{lem:case} and Assumption~\ref{assum:zero_zero}, $f^{\tm,s}_k$ converges to one almost surely.
Denote the numerator and the denominator of $f^{\tm,s}_k$ by $f^{\tm,s}_{{\rm N},k}$ and $f^{\tm,s}_{{\rm D},k}$.
Since $0<f^{\tm,s}_{{\rm D},k}\leq1$, we have $0\leq f^{\tm,s}_{{\rm D},k}|f^{\tm,s}_{k}-1|\leq |f^{\tm,s}_{k}-1|.$
Because $|f^{\tm,s}_{k}-1|\to 0$ almost surely and $f^{\tm,s}_{{\rm D},k}$ is bounded, we have $f^{\tm,s}_{{\rm D},k}|f^{\tm,s}_{k}-1|\to 0$ almost surely.
This leads to that $|p^{\htheta_{\rm b},\tm,s}_k-p^{\htheta_{\rm m},\tm,s}_k|(1-\pi^{\htheta=\tm,}_k)\to 0$
almost surely.
Since $s$ is a Bayesian-Nash equilibrium with detection-averse utilities, the claim holds.
\end{proof}

\vspace{3mm}
\begin{proof}
\emph{Proof of Theorem~\ref{thm:asymptotic_security}:}
First, note that the claim is equivalent to
\[
 {\rm Pr}(\underbrace{\{A^{\htheta_{\rm b},\tm,s}_k \neq  A^{\htheta_{\rm m},\tm,s}_k\ \io\}}_{=:E})=0
\]
from the finiteness of $\mc{A}$.

Because the Markov decision process is finite, Lemma~\ref{lem:pp} is equivalent to
\[
 {\rm Pr}(\{ \underbrace{p^{\htheta_{\rm b},\tm,s}_k \neq p^{\htheta_{\rm m},\tm,s}_k}_{=:F_k}\ \io \})=0.
\]
This is equivalent to ${\rm Pr}(F)=0$ where
\[
 F:= \left\{
 \omega \in \Omega:
 \sum_{k=1}^\infty {\rm Pr}(F_k|\sigma(X^{\tm,s}_{1:k-1}))(\omega)=\infty
 \right\}
\]
from the generalized second Borel-Cantelli lemma\cite[Theorem~4.3.4]{Durrett2019Probability}.
Now assume ${\rm Pr}(E)\neq0$.
From ${\rm Pr}(E\cap F)=0$ and ${\rm Pr}(E\cap F)={\rm Pr}(F|E){\rm Pr}(E)=0,$
we have ${\rm Pr}(F|E)=0$.
We here show ${\rm Pr}(F|E)\neq0$ and prove the claim by contradiction.

Take $\omega\in E$.
Then there exists a subsequence $\{k_i\}_{i\in\NN}$, which depends on $\omega$, such that $A^{\htheta_{\rm b},\tm,s}_{k_i(\omega)} \neq  A^{\htheta_{\rm m},\tm,s}_{k_i(\omega)}$ holds for any $i\in\NN$.
For this subsequence,
\[
 \sum_{k=1}^\infty {\rm Pr}(F_k|\sigma(X^{\tm,s}_{1:k-1}))(\omega) \geq \sum_{i=1}^\infty {\rm Pr}(F_{k_i(\omega)}|\sigma(X^{\tm,s}_{1:k_{i-1}(\omega)}))(\omega)
\]
holds.
From the finiteness of the Markov decision process and Assumption~\ref{assum:input_obs}, we have
\[
 \inf_{i\in\NN} {\rm Pr}(F_{k_i(\omega)}|\sigma(X^{\tm,s}_{1:k_{i-1}(\omega)})) (\omega)>0.
\]
Thus $\sum_{k=1}^\infty {\rm Pr}(F_k|\sigma(X^{\tm,s}_{1:k-1})) (\omega)=\infty$
holds for $\omega\in E$.
Therefore ${\rm Pr}(F|E)=1$, which leads to a contradiction.
\end{proof}

\bibliographystyle{IEEEtran}
\bibliography{sshrrefs}

\end{document}